\documentclass[twocolumn]{aastex631}
\usepackage{amsmath}
\usepackage{graphicx}
\usepackage{amsmath}
\usepackage{bm}
\usepackage{hyperref}
\renewcommand{\figurename}{Fig}
\begin{document}
	
\title{Interpretations of the $10\%$ polarization observed in the early forward-shock afterglow of GRB 091208B}
	
\author{Yang liu}
\author[0000-0001-5641-2598]{Mi-Xiang Lan}
\affiliation{Center for Theoretical Physics and College of Physics, Jilin University, Changchun 130012, People's Republic of China; lanmixiang@jlu.edu.cn}


\begin{abstract}
The $\sim10\%$ optical polarization observed at the early stage of GRB 091208B comes from the forward shock emission, which is higher than the conventionally predicted value. Polarizations of the forward shock radiation would depend on the observational geometry and the post-shock magnetic field structure. This magnetic field could arise either from the compression of a pre-existing magnetic field (i.e., the magnetic field in the outer medium) or from the shock-generated instabilities. In this paper, we use a synchrotron radiation model to fit the light curve and polarization observations of GRB 091208B. Two scenarios are considered: one is the case of a slightly off-axis observer, and the other is with a large-scale ordered magnetic field component in the burst environment. We found both scenarios could interpret the observations of GRB 091208B. For the slightly off-axis observation scenario, the observational angle is restricted to be within the range of (1.02, 1.05) times the jet half-opening angle. For the large-scale ordered magnetic field component scenario, the ratio between the ordered component to the random component is constrained to be around 1.

\end{abstract}


\section{INTRODUCTION}
Gamma-ray bursts (GRBs) are brief high-energy radiation bursts that randomly occur in the sky (\citealp{1995ARA&A..33..415F}).
These bursts have extremely diverse durations, temporal characteristics and spectral variations (\citealp{1995ARA&A..33..415F}).
Their spatial distribution is isotropic but shows radially inhomogeneous (\citealp{1995ARA&A..33..415F}).
The emission of the GRBs include two stages : the prompt emission phase and the afterglow phase.
The afterglow is produced by the interaction of the relativistic jets with the surrounding medium, spanning from the radio to gamma-ray bands.
It gradually decays over a time scale from several hours to several days (for reviews, see \citealp{1999PhR...314..575P,2000ARA&A..38..379V,2002ARA&A..40..137M,2004IJMPA..19.2385Z}).
Since the first optical afterglow was discovered in GRB 970228 (\citealp{1997IAUC.6584....1G,1997Natur.386..686V}), a lot of optical afterglows have been detected (for a summary, see \citealp{2022AAS...24010906D}).
Meanwhile, many of these afterglows were detected with polarizations (\citealp{2016A&AT...29..205C}).
Moreover, the afterglow can locate the burst source and the host galaxy, reveal the environment around the burst, and be used as a cosmological probe (\citealp{2001ApJ...562..654D,2009ApJS..185..526F}).

GRB 091208B is the first GRB afterglow observed to have optical polarization phenomenon in the early forward shock radiation (\citealp{2012ApJ...752L...6U}). The optical polarization of this burst was detected from time $t = 149$ s to $t=706$ s, with an averaged polarization degree (PD) of $10\%\pm2.5\%$. Its redshift is 1.063 and it is a long burst with a duration of $\sim14.9$ s (\citealp{2012ApJ...752L...6U}). The afterglow emission, as a point source, is commonly thought to be originated from the synchrotron radiation after the shock. Therefore, its polarization depends strongly on the magnetic field structure and the observational geometry (\citealp{1999ApJ...524L..43S,1999ApJ...525L..29G,2003Natur.423..388W,2003ApJ...594L..83G,2009ApJ...698.1042T,2012ApJ...752L...6U,2019ApJ...870...96L,2023ApJ...943..118K,2024ApJ...977..224K}). 

It has been commonly believed that the magnetic field in the afterglow emission region is produced by the instabilities in the collisionless shocks (\citealp{1999ApJ...526..697M,2007ApJ...671.1858S}). However, the time-averaged PD of GRB 091208B is greater than that produced by forward shock with a random magnetic field generated by two-stream instability observed on-axis (\citealp{1999ApJ...524L..43S,1999ApJ...525L..29G,2016ApJ...816...73L,2023ApJ...952...31L}). \cite{2012ApJ...752L...6U} suggested that the magnetic patchy with larger coherent scale generated by magnetohydrodynamic instability may give rise to the observed high PD. However, \cite{2024ApJ...977..224K} demonstrated that magnetic patchy model could not be account for such high PD of GRB 091208B. Later, \cite{2021MNRAS.505.2662J} suggested the high PD of the burst is due to the contributions from the highly polarized reverse shock emission. Although the total flux density is dominated by the forward shock region, the polarized flux can be dominated by the reverse shock region.

Except the field generated by instabilities, the compressed medium magnetic field can not be ignored and can play a key role in the events with low density and low energy participation factor of magnetic field \citep{2010MNRAS.409..226K}, such as GRB 170817 (\citealp{2021MNRAS.507.5340T}). The environment around the GRB may contains the ordered field. Such ordered magnetic field may originate from the stellar wind of the GRB progenitor, for example a Wolf-Rayet (WR) star \citep{2016MNRAS.458.3381H}. Globally, the magnetic field in the wind is primarily toroidal, but it can be considered as aligned locally \citep{2006NJPh....8..131L}.

In this paper, we use the model of the synchrotron radiation from the forward shock in a medium with a power-law number density to explain the polarization phenomenon of the afterglow of GRB 091208B. Two scenarios are considered: the slightly off-axis observation with a random field generated by the two-stream instability and the forward shock with an ordered magnetic field component in the stellar wind environment. The structure of the paper is as follows: in Section \ref{2}, we briefly introduce the model; in Section \ref{3}, we present our numerical results; in Section \ref{4}, we give the conclusions and discussion.

\section{The model}\label{2}

A simple synchrotron radiation model is adopted here: the extremely relativistic jet ejected by the central engine collides with the outer medium, generating a forward shock that propagates into the medium. In this study, a power-law density profile of the outer medium  $n(r)=n_0(r/r_0)^{-k}$ is adopted. The parameter $n_0$ is the normalization of $n(r)$ at radius $r_0$. The model does not take into account the lateral expansion effect.

The forward shock radiation originates from the synchrotron radiation of the relativistic electrons, and its flux and polarizations depend on the strength and configuration of the magnetic field in the radiation region. The magnetic field behind the shock may be created by two mechanisms: one is the compression of the existing magnetic field in the outer medium \citep{1980MNRAS.193..439L}, the other is the  instabilities generated by the shock, e.g., the Weibel instability (\citealp{1999ApJ...526..697M}).

Due to the compression caused by forward shock, the non-radial components of the background magnetic field within the medium would be amplified (\citealp{2000ApJ...534..239M}). The magnetic field strength in the compressed medium is $B_{comp} = 4 \Gamma B_{M}$, where $\Gamma$ is the Lorentz factor of the shock and $B_M$ is the background field strength (\citealp{2021MNRAS.507.5340T}). If the burst occurs in the stellar wind of a WR star, the magnetic field in the shocked region of such an environment would be a mixed field with the ordered aligned component being amplified by the shock, as well as the random component generated by the instabilities. The typical strength of the surface magnetic field of the WR star is around 1000 G \citep{2016MNRAS.458.3381H}, so the field strength at the emission radius of the afterglow is about 0.5 mG according to the conservation of the magnetic flux and it is roughly 0.2 G in the shock compressed region. And the inferred field strength from the following fitting of the light curves around the time of the polarization detection is $\sim$0.15 G, which is consistent with this scenario.

For the polarization model, we adopt \cite{2016ApJ...816...73L, 2023ApJ...952...31L}. The detailed formula can be referred to the above mentioned references. Since the Stokes parameter $U_\nu$ is zero, the time resolved PD of the radiation from a jet with a random field in the radiation region can be expressed as follows.
\begin{equation}
    \label{equation1}
	PD=\frac{Q_\nu}{F_\nu}.
\end{equation}
where $F_\nu$ represents the flux density and $Q_\nu$ is the Stokes parameter Q. When the Stokes parameter $Q_\nu$ changes its sign, the polarization angle (PA) would change abruptly by $90^\circ$.

The time-resolve PD and PA of a jet with an aligned magnetic field in the emission region can be expressed as follows.
\begin{equation}
    \label{equation1}
	PD=\frac{\sqrt{Q_\nu^2+U_\nu^2}}{F_\nu}.
\end{equation}
\begin{equation}
    \label{equation2}
	PA=\frac{1}{2} arctan\left(\frac{U_\nu}{Q_\nu}\right).
\end{equation}

The local PD in an aligned field is expressed as (\citealp{2023ApJ...952...31L})
$$
\pi_0=\frac{\int G(x) N\left(\gamma_e\right) d \gamma_e}{\int F(x) N\left(\gamma_e\right) d \gamma_e}
$$
where $F(x)=x \int_x^{\infty} K_{5 / 3}(t) dt$ and $G(x)=x K_{2 / 3}(x)$. And $x=\nu^{\prime} / \nu^{\prime}_c,\ \nu^{\prime}=(1+z) \nu / \mathcal{D}$ and $\nu^{\prime}_c$ are the observational frequency in the comoving frame and the critical frequency of electrons with Lorentz factor $\gamma_e$, respectively. $K_{5 / 3}(x)$ and $K_{2 / 3}(x)$ denote the modified Bessel functions with $5 / 3$ and $2 / 3$ orders.

In order to consider the EATS effect, it is convenient to express the dynamical quantities (such as the bulk Lorentz factor $\gamma$ of the shocked region) as a function of the radius $r$. The EATS of the afterglow emission is (\citealp{1998ApJ...494L..49S}):
\begin{equation}
    \label{equation3}
	t_b-\frac{rcos\theta}{c}=\frac{t}{1+z}.
    \end{equation}
where $t_b$ represents the dynamic time in the burst source frame, $\theta$ represents the angle between the local velocity direction and the line of sight, $c$ is the speed of light, and $z$ is the redshift of the source.

\section{Numerical Results}\label{3}

The $\sim10\%$ optical polarization observation of GRB 091208B was performed from 149 s to 706 s, of which the afterglow emission of the burst  was fitted by the forward shock radiation \citep{2012ApJ...752L...6U}. High polarization observed in the early afterglow of GRB 091208B may be attributed to four possible causes: polarization from turbulent magnetic fields amplified by magnetohydrodynamic instabilities, the contribution of a reverse shock, slightly off-axis observation, and the presence of a large-scale ordered magnetic field in the burst environment. Since the first two possibilities had been considered in the former studies \citep{2012ApJ...752L...6U,2021MNRAS.505.2662J,2024ApJ...977..224K}, the last two scenarios are considered here.

\begin{figure}
\centering
\includegraphics[scale=0.4]{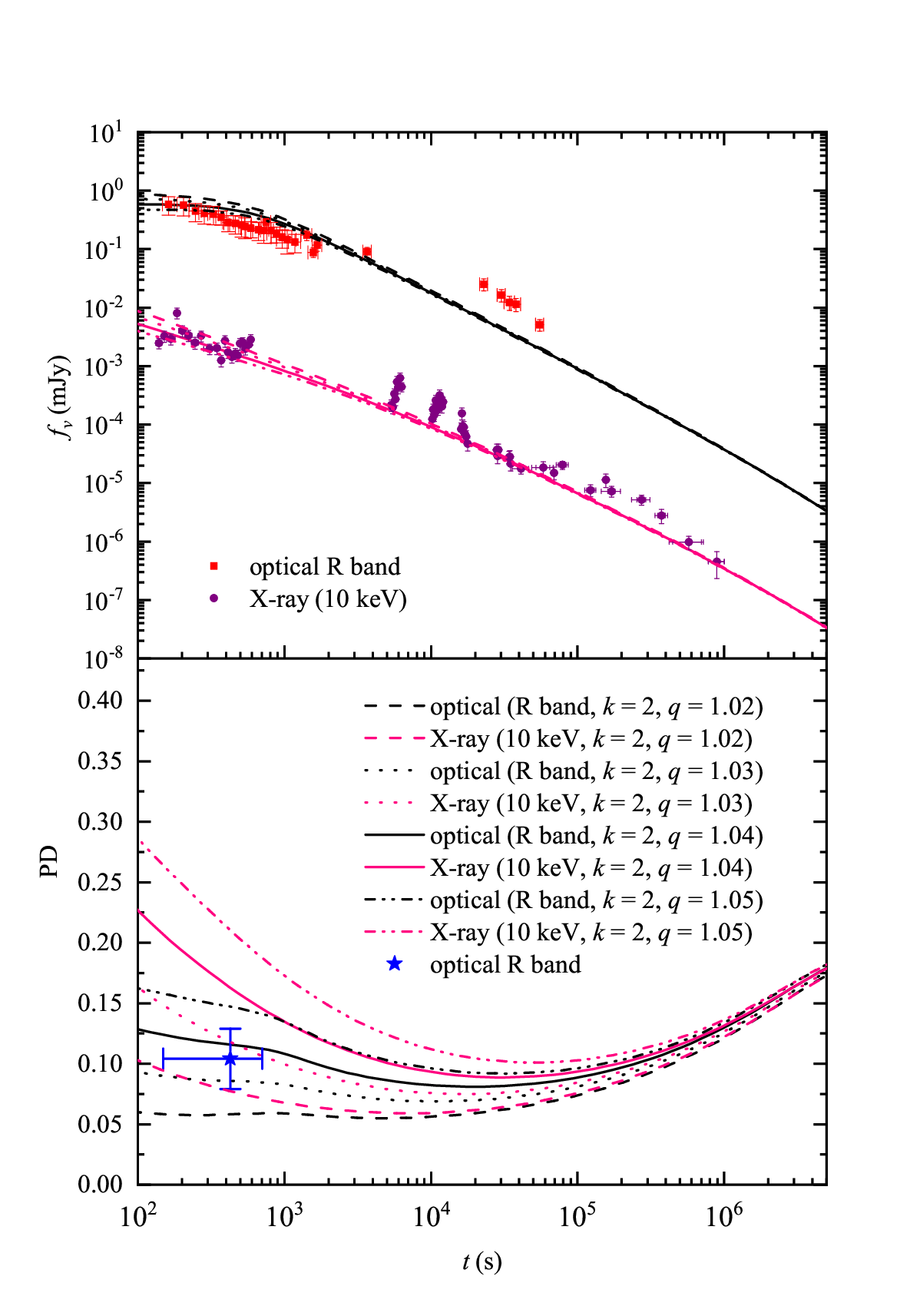}
\caption{Fitting of the observational data of GRB 091208B with a slightly off-axis observational geometry. The upper and lower panels show the light curves and the PD curves, respectively. The black lines (optical R band) and pink lines (10 keV) represent the predicted curves. The red squares (purple circles) show the observational data in optical R band (X-ray 10 keV). And the blue star represent the observed time-averaged PD from 149 s to 706 s in optical R band.}
\label{pianzhou}
\end{figure}

For the off-axis observation scenario, the light curves and the optical polarization of GRB 091208B can not be fitted simultaneously in an uniform outer medium with a random field. So a wind environment should be involved, which is also consistent with the collapsar progenitor for the long GRBs (as for GRB 091208B). And the off-axis observation in a stellar wind environment ($k=2$) with a shock-generated random field would be consistent with the observational data when $q\equiv{\theta_V/\theta_j}$ is restricted to be within the range of (1.02, 1.05). The $\theta_V$ is the observational angle, which is the angle between the jet axis and the line of sight. The $\theta_j$ is the half-opening angle of the jet. The other parameters used in the fitting are as follows: the isotropic equivalent energy $E_{52}=E_{iso}/(10^{52} \ erg)$$=8$, the initial Lorentz factor $\eta=200$, the initial width of the jet $\Delta_0=10^{10}$ cm, $\theta_j=0.1$ rad, $n_0=1$ $cm^{-3}$, $r_0=10^{17}$ cm, $\varepsilon_e=0.3,\  \varepsilon_B=3 \times 10^{-4}$ and $p=2.07$. The $\varepsilon_e$ and $\varepsilon_B$ are the energy participation factor of the electrons and the magnetic field, respectively. The spectral index of the injected electrons is denoted as $p$. The shock is assume to be adiabatic. The optical data is taken from \citealp{2012ApJ...752L...6U} and the X-ray data is from the $Swift$ website \footnote{\url{https://www.swift.ac.uk/}} and \cite{2007A&A...469..379E,2009MNRAS.397.1177E}. The corresponding fitting results are shown in Fig \ref{pianzhou}. We can see that the polarization is lower than the observed value for $q=1.02$ and is higher for $q=1.05$.

\begin{figure}
\centering
\includegraphics[scale=0.4]{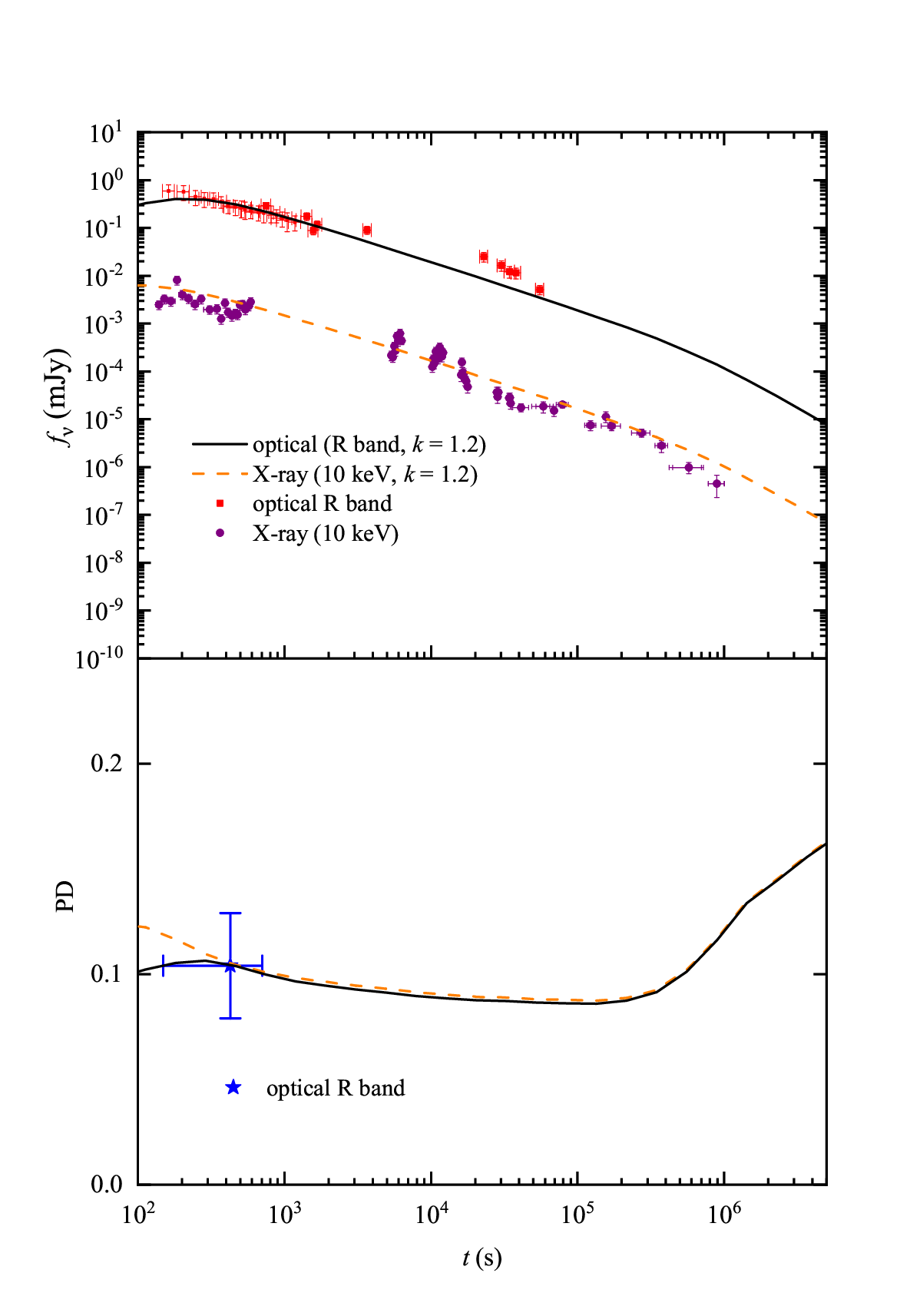}
\caption{Fitting of the observational data of GRB 091208B with a mixed magnetic field in the forward shock region. The upper and lower panels show the light curves and the PD curves, respectively. The red squares represent the data in optical R band, while the purple circles represent that in 10 keV X-ray band. The blue star is the observed time-averaged PD from 149 s to 706 s in optical band. The solid (optical R band) and dashed (10 keV) lines represent the fitting results.}
\label{guangbian}
\end{figure}

For the scenario with a large-scale ordered field component in the wind environment, the parameters we take in our fitting are: $E_{53}=E_{iso}/(10^{53} \ erg)=4$, $\eta=200$, $\Delta_0=10^{10}$ cm, $\theta_j=0.1$ rad, $n_0=1$ $cm^{-3}$, $r_0=10^{17}$ cm, $k=1.2$, $\varepsilon_e=0.2, \varepsilon_B=10^{-5}$, $p=2.08$, and the field orientation $\delta = \pi/4$. The shock is assume to be adiabatic. The predicted upper limit of the time-averaged optical PD from $t=149 $ s to $t=706$ s is $50.8\%$ with a large-scale ordered aligned field in the wind environment, which would be higher than the observed $\sim10\%$ PD. Hence the field in the outer medium should be mixed. By varying the parameter $\xi_B$, which represents the ratio of the ordered to the random component of the magnetic field \citep{2019ApJ...870...96L,2021MNRAS.507.5340T}, we find that $\xi_B=1$ for on-axis observation ($q=0$) would be consistent with the observational data, as shown in Fig \ref{guangbian}.

To examine how the polarizations would depend on the values of $\xi_B$ and the field orientation $\delta$, the light curves and the corresponding polarization curves in the wind environment are calculated by varying the two parameters. The other parameters are set as follows: $E_{52}=E_{iso}/(10^{52} \ erg)=6$, $\eta=200$, $\Delta_0=10^{10}$ cm, $\theta_j=0.1$ rad, $n_0=1$ $cm^{-3}$, $r_0=10^{17}$ cm, $\varepsilon_e=0.3,\  \varepsilon_B=3 \times 10^{-5}$ and $p=2.1$. And we set $\xi_B=1$ and vary $\delta$. Both the light curve and the PD curve show almost no dependence on $\delta$ value. The time-resolved PA is a constant with time and its value would depend on the concrete  $\delta$ value. For two arbitrary PA curves, the difference of their time-resolved PA values would be same as the difference for the parameter  $\delta$. Then we set $\delta = \pi/4$ and vary $\xi_B$. The light curve is almost independent on the value of $\xi_B$. In contrast, the time-resolved PD values would increase with $\xi_B$ as expected. The time-resolved PA is also roughly a constant with time\footnote{The change of the PA value during the whole afterglow stage would be less than $10^\circ$.} and its value would depend on the value of $\xi_B$.

\begin{figure*}
\centering
\includegraphics[scale=0.4]{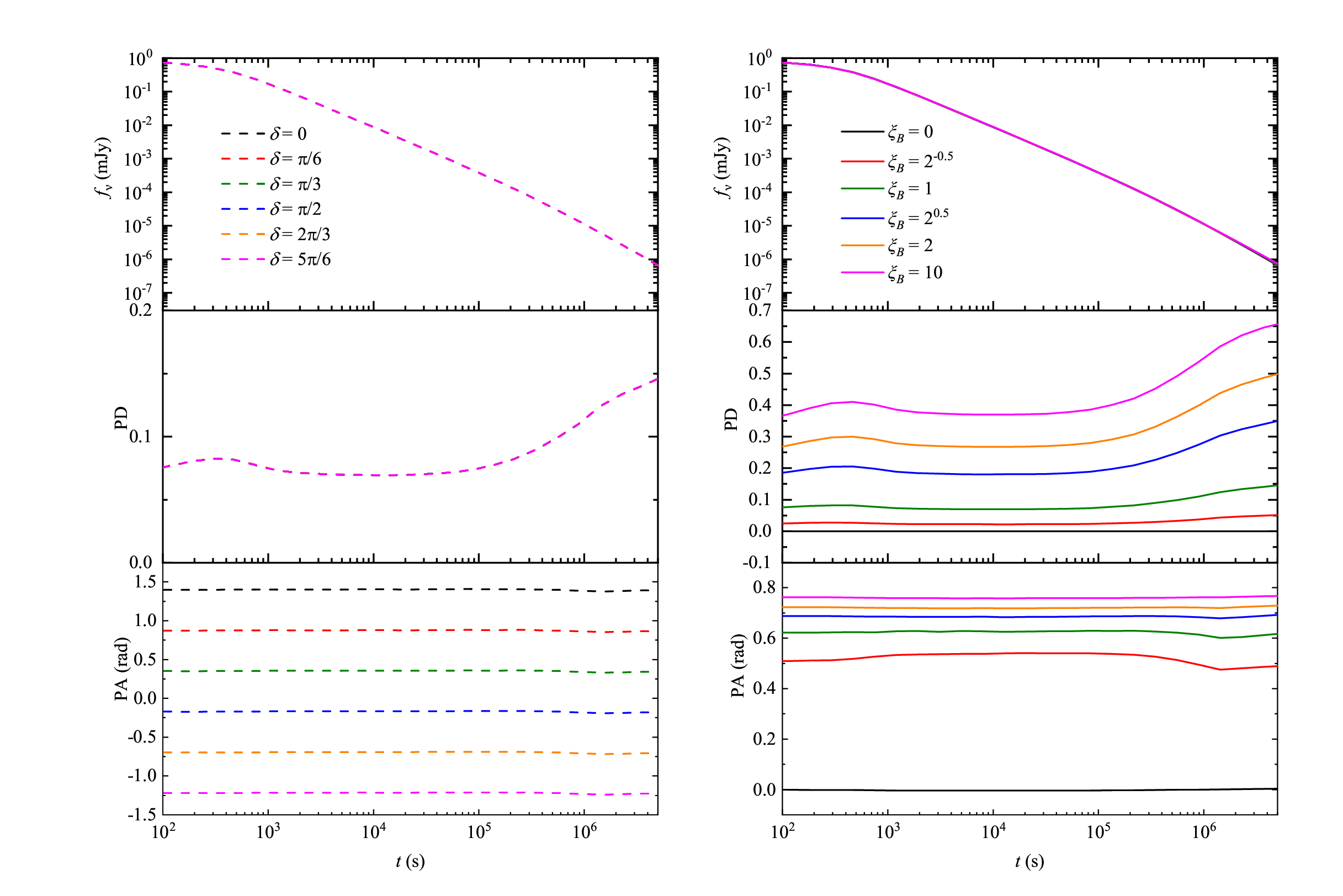}
\caption{The impacts of the magnetic field parameters $\delta$ (left) and $\xi_B$ (right) on the light curves and polarization curves. The upper panel shows the light curves. The middle and lower panels show the PD and PA evolutions, respectively.}
\label{bili}
\end{figure*}

\section{Conclusions and Discussion}\label{4}

Polarization observation from early forward shock emission of GRB 091208B is $\sim10\%$ \citep{2012ApJ...752L...6U}, which seems to be curious. The predicted optical polarizations of GRB afterglow from the forward shock radiation would be at late time around the the jet break and usually smaller than $\sim10\%$ \citep{1999ApJ...524L..43S,1999ApJ...525L..29G,2016ApJ...816...73L, 2023ApJ...952...31L}. And the polarization observations of GRB optical afterglow show that PDs would be less $\sim5\%$ after 1000 s and less than $2\%$ after $10^5$ s (for a review, see \citealp{2016A&AT...29..205C,2022arXiv220803583L}), which is consistent with the scenario that the afterglow emission is generated by the synchrotron radiation from the forward shock region within a random field observed on-axis. Therefore, a $10\%$ optical PD was relatively high for the early-time forward shock emission.

In this paper, we use two scenarios to interpret such an observation, the slightly off-axis observation and large-scale ordered magnetic field component in the forward shock region. And both of the two scenarios could interpret the observations (including the light curves and polarization) of the burst. The observational angle is limited to the range $(1.02 < q < 1.05)$ for the slightly off-axis observation scenario. Since the observational angle in the limited range is still within the $\theta_j+1/\eta$ cone, the prompt emission of the burst is visible, which is consistent with the observation. The parameters we take are reasonable and the model predictions could interpret the most observations of the burst.

For the scenario with a large-scale field component in the outer medium, the predicted upper limit of PD is roughly ($40-50$)$\%$ and the indicated ratio between the ordered component and random component is about 1. Such an ordered magnetic field component in the medium is not rare, since there are many interstellar environments that would have a large-scale ordered field component, e.g., in the the stellar wind of a WR star \citep{2016MNRAS.458.3381H}. Therefore, polarizations can be used as a probe of magnetic field structure in the environment of GRBs. Since we can not exclude the slightly off-axis scenario, the mixed field scenario would only be another possibility.

The dependence of the time-resolved polarizations on the parameters of the magnetic field are also investigated. As expected, the predicted time-resolved PD would depend strongly on the ratio between the ordered to random component ($\xi_B$) and would increase with $\xi_B$. However, the time-resolved PD would independent on the field orientation ($\delta$). The predicted time-resolved PA is roughly a constant with observational time and its concrete value would depend on both the ratio and the field orientation.

\begin{acknowledgments}
We thank the anonymous referee for useful comments that improved our manuscript.
This work is supported by the National Natural Science Foundation of China (grant No. 12473040).
M.X.L. would also like to acknowledge the financial support from Jilin University.
\end{acknowledgments}

\twocolumngrid

\bibliography{reference}
\end{document}